\documentclass[prc,twoside,superscriptaddress,showkeys,showpacs,twocolumn,nofootinbib]{revtex4}

\usepackage{graphicx, latexsym, amssymb, amsmath, color, multirow, mathrsfs, CJK, ifpdf}

\usepackage[section]{placeins}
\usepackage{amsmath}
\usepackage{amsfonts}
\usepackage{amssymb}
\usepackage{bm}
\usepackage{graphicx}
\usepackage{color} %
\usepackage{txfonts}
\makeatletter

\newcommand{\Rmnum}[1]{\expandafter\@slowromancap\romannumeral#1@}
\makeatother

\begin{document}


\title{Shape vibrations and quasiparticle excitations in the lowest  $0^{+}$ excited state of the Erbium isotopes}\author{Fang-Qi Chen}
\author{J. Luis Egido}
\email{j.luis.egido@uam.es}
\affiliation{Departamento de F\'isica Te\'orica, Universidad Aut\'onoma de Madrid, E-28049, Madrid, Spain}

\date{\today}
\begin{abstract}
The ground and first  excited $0^{+}$ states of the  $^{156-172}$Er isotopes are analyzed in the framework of the generator coordinate method.
The shape parameter $\beta$ is used to generate wave functions with different deformations which together with the  two-quasiparticle states built on them provide a set of states. An angular momentum and particle number projection 
of the latter spawn  the basis states of the generator coordinate method. 
With this ansatz  and using the separable pairing plus quadrupole interaction we obtain a good agreement with the experimental spectra and E2 transition rates up to moderate spin values. The structure of the wave functions suggests that the first  excited $0^{+}$ states in the soft Er isotopes are dominated by shape fluctuations, while in the well deformed Er isotopes the two-quasiparticle states are more relevant.  In between both degrees of freedom are necessary .
\end{abstract}

\maketitle

\section{Introduction}\label{sect1}
The nature of the lowest-lying $0^{+}$ excited states (denoted as $0^{+}_{2}$ state in the following) in deformed nuclei has been a long standing problem in nuclear physics and studied by various approaches  \cite{Garrett}. Traditionally they have been considered to be collective excitations such as the $\beta$-vibration\cite{BM.88}. In recent years there has been calculations along this line based  based on the algebraic collective model of Rowe and co-workers \cite{Rowe} as well as  analytical solutions of the Bohr Hamiltonian  with a certain kind of potential and a deformation-dependent mass term \cite{Bohr2011,Bohr2013}. There are also calculations using the interacting boson model (IBM) with different truncated Hamiltonians \cite{IBM1997,IBM2004,IBM2011,IBM2012}. In these calculations the  $0^{+}_{2}$ state is supposed to be a pure collective excitation, and its excitation energy can be fitted together with the Yrast and the $\gamma$-bands.

 Other studies do not assume  that the $0^{+}_2$ states  are purely collective. For example, 
in the quasiparticle-phonon model (QPM)   the $0^{+}_2$ excitations have been investigated in several rare-earth nuclei \cite{QPM-Er166,QPM-Er168,QPM-Gd158}. In these works the first $0^{+}_2$ excitations are described as one-phonon states, with the  result that the phonons are built by several (not many) two-quasiparticle configurations.  A similar conclusion is also obtained in the studies of the $0^{+}_2$ excitations in $^{158}$Gd and $^{168}$Er with the projected shell model (PSM) \cite{PSM-Gd158,PSM-Er168}. These studies reveal the relevance of the two-quasiparticle states in the $0^{+}_{2}$ states.  However, the shortcoming of the latter models is that they do not work well in transitional nuclei. Recently, there has been also calculations of the
low-lying excited states with the Quasi-Random Phase Approximation with Skyrme forces \cite{Tera,HK.13} and the 
five dimensional collective Bohr Hamiltonian with the Gogny force \cite{Bohr_5DM}.

In the treatment of transitional nuclei, the generator coordinate method (GCM) is widely used  \cite{Ring}. The GCM is a very flexible microscopic model  which handles shape vibrations as the Bohr Hamiltonian but that at the same time allows to incorporate single particle degrees of freedom. It has been applied with great success with effective forces,  like Skyrme  \cite{Skyrme}, Gogny  \cite{Gogny} or relativistic  \cite{Relativistic},  to consider situations where the collective degrees of freedom play a relevant role, like shape coexistence in transitional nuclei.  In recent years  the GCM has been used also with separable forces to calculate the first $0^{+}$ excitations in several Gd, Dy and Er isotopes (including the transitional $N=90$ isotones) \cite{GCM2013}. In this work the quadrupole deformation $\beta$ is used as the generator coordinate and after projecting on angular momentum and particle number the Hill-Wheeler equation \cite{HW} is solved.  In these calculations the $0^{+}_2$  states are interpreted as shape vibrations. The calculated excitation energies and transition probabilities of the $0^{+}_{2}$ states are in good agreement with the experimental data. However, this study is restricted to isotopes with neutron number $N\leq98$.   An attempt to extend this study to the heavier Er isotopes with $N>98$  was not successful.

Considering the above mentioned studies with the QPM and PSM, one may guess that the failure of the GCM calculation in heavier Er isotopes is due to the lack of two-quasiparticle states in the model space. With the inclusion of the two-particle states, one could at least expect results as good as the PSM calculations for the heavier Er isotopes with $N>98$. In fact, there are several hints indicating that the first $0^{+}$ excitations in these Er isotopes are not as collective as a pure shape vibration. A measurement of the lifetime of the $0^{+}_{2}$ excitation in $^{166}$Er \cite{Er166BE2} showed that the transition probability of the $0^{+}_{2}$ state in this nucleus is not as large as expected for a $\beta$-vibration. The best candidate for the $\beta$-vibration in this nucleus is the third $0^{+}$ excited state. In $^{168}$Er, the measured E2 transition probability  from the $0^{+}_{2}$  to the $2^{+}_{1}$ state is also very small \cite{Er168BE2}. The $B(E2)$ value of $0.08(1)$ W.u. is much smaller than the value of tens of W.u. expected for a $\beta$-vibration \cite{Garrett}.

 There are also some hints in this direction from the above mentioned calculations with the Bohr model and the  IBM. In the calculations with the Bohr model, the parameters are determined by fitting the experimental spectra, and the transition rates are then calculated with these parameters. It turns out that the $B(E2)$ rates from the $0^{+}_2$ states are always overestimated, sometimes by one order of magnitude. This indicates that the calculated $0^{+}_{2}$ states are "more collective" than they should be. In the IBM calculations the parameters are also determined by fitting the experimental data. It was found in Ref.~\cite{IBM2004} that the parameters determined in this way for $^{168}$Er do not follow the smooth trend  as a function of the neutron number. However, if one assumes that the lowest collective $0^{+}$ excitation is the second $0^{+}$ excitation in $^{168}$Er, one would get smoothly varying parameters. This also indicates that the first $0^{+}$ excitation in $^{168}$Er may not be a collective $\beta$-vibration.

The purpose of this paper is to generalize  the above mentioned GCM calculation \cite{GCM2013}  by simultaneously considering different nuclear shapes and their two-quasiparticle excitations  in the GCM ansatz. We apply the new theory to perform a systematic study of the first $0^{+}$ excitations in the Er isotopes from $N=88$ to $N=104$. Since the light isotopes are very soft in $\beta$ and the heavy ones strongly deformed, it is expected that the role played by the collective and the single particle degrees of freedom as well as their coupling will be elucidated. It is expected that the inclusion of the two-quasiparticle states   
will improve  the description of these $0^{+}$ states in the Er isotopes with larger neutron numbers. It should be mentioned that the GCM calculation with two-quasiparticle states has been used in Ref.~\cite{GCM+QP1976,GCM+QP1977} in a very restricted configuration space for the study of Ge and Zn isotopes. 

In Sect.\ref{sect2} we give a presentation of the theoretical framework and some numerical details. The results are shown and discussed in Sect.\ref{sect3}. We give a summary of our conclusion in Sect.\ref{sect4}.

\section{Theory and model space}\label{sect2}

  In our calculations we use the separable pairing plus quadrupole Hamiltonian  \cite{GCM2013,PSMreview}
\begin{equation}\label{Hamiltonian}
\hat{H}=\hat{H}_{0}-\frac{1}{2}\sum_{\tau \tau'}\chi_{\tau \tau'}\sum_{\mu}\hat{Q}^{\dag}_{\tau\mu}\hat{Q}_{\tau'\mu}-G_{M}\sum_{\tau}\hat{P}^{\dag}_{\tau}\hat{P}_{\tau}-G_{Q}\sum_{\tau \mu}\hat{P}^{\dag}_{\tau\mu}\hat{P}_{\tau\mu},
\end{equation}
where the operators $\hat{H}_{0}$, $\hat{Q}_{\mu}$, $\hat{P}$ and $\hat{P}_{\mu}$  ( $\pi$ for protons and  $\nu$ for neutrons) are given by:
\begin{eqnarray}
\hat{H}_{0}= \sum_{k} \epsilon_{k}c^{\dag}_{k}c_{k},  \;\;\; \hat{Q}_{\mu}=\sum_{k,l}(Q_{\mu})_{kl}c^{\dag}_{k}c_{l},\\
\hat{P}=\frac{1}{2}\sum_{k}c_{k}c_{\bar{k}}, \;\;\; \hat{P}_{\mu}=\frac{1}{2}\sum_{k,l}(Q_{\mu})_{kl}c_{k}c_{\bar{l}}.
\end{eqnarray}
with the quadrupole operator defined by  $\hat{Q}_{\mu} =  r^2  Y_{2\mu}$
and $\mu$ runs from $-2$ to $2$.

 As mentioned in the introduction as an ansatz for our wave functions we use the shape parameter $\beta$ to
 generate wave functions $|\Phi_{0}(\beta)\rangle$  with this shape.  For this purpose we solve the  Hartree-Fock-Bogoliubov  equation with constraints on the total quadrupole moment $\hat{Q}_0$ and the average particle number. The
 wave function  $|\Phi_{0}(\beta)\rangle$ of the energy minimum for a given $\beta$ value is provided by
\begin{equation}
\delta\langle\Phi_{0}(\beta)|\hat{H}-\lambda_{n}\hat{N}-\lambda_{p}\hat{Z}-\lambda_{q}\hat{Q}_{0}|\Phi_{0}(\beta)\rangle=0,
\label{HFB_Eq}
\end{equation}
with the Lagrange multipliers $\lambda_{n}$, $\lambda_{p}$ and $\lambda_{q}$  determined by the constraining conditions:
\begin{eqnarray}
\langle\Phi_{0}(\beta)|\hat{N}|\Phi_{0}(\beta)\rangle=N, \nonumber \\
 \langle\Phi_{0}(\beta)|\hat{Z}|\Phi_{0}(\beta)\rangle=Z, \\
\langle\Phi_{0}(\beta)|\hat{Q}_{0}|\Phi_{0}(\beta)\rangle=Q_{0}.\nonumber
\end{eqnarray}
The relation between the quadrupole operator and the deformation parameter $\beta$ is given by  $Q_{0}=\sqrt{\frac{16\pi}{5}}\frac{3}{4\pi}ZeR^{2}_{0}\beta$ with $R_{0}=r_{0}A^{1/3}$ and $r_{0}=1.2$ fm. For each HFB vacuum  $|\Phi_{0}(\beta)\rangle$ there is a set of corresponding quasiparticle operators $\alpha_i(\beta)$ satisfying
\begin{equation}
\alpha_i(\beta)|\Phi_{0}(\beta)\rangle =0,   \;\; \forall i.
\end{equation}
The second components of our GCM ansatz are the two-quasiparticle states, defined by 
\begin{equation}
|\Phi_{ij}(\beta)\rangle=\alpha^{\dag}_{i}(\beta)\alpha^{\dag}_{j}(\beta)|\Phi_{0}(\beta)\rangle.
\end{equation}
Finally,  the HFB vacua and the two-quasiparticle states are projected onto good angular momentum and particle number. Thus
 the complete ansatz for our wave function has the form
\begin{eqnarray}\label{eqwf}
|\sigma,IM \rangle&=&\int d\beta f^{\sigma,I}_{0}(\beta) \hat{P}^{I}_{MK}\hat{P}^{N}\hat{P}^{Z} |\Phi_{0}(\beta)\rangle\nonumber\\
& & \;\;\;\;+\sum_{ij}\int d\beta f^{\sigma,I}_{ij}(\beta) \hat{P}^{I}_{MK}\hat{P}^{N}\hat{P}^{Z} |\Phi_{ij}(\beta)\rangle\nonumber \\
& = & \sum_{\rho} \int d\beta f^{\sigma,I}_{\rho}(\beta) \hat{P}^{I}_{MK}\hat{P}^{N}\hat{P}^{Z} |\Phi_{\rho}(\beta)\rangle,
\end{eqnarray}
where the index $\rho$ runs over the set $\{0, (ij)\}$ and $\sigma$ labels the different states with angular momentum $I$.
In this calculation the axial symmetry is preserved, and each of the HFB vacua and the two-quasiparticle states has a good quantum number $K$. Therefore the summation over $K$ is omitted in Eq.~\ref{eqwf}. The projection operators in the above expression are given by \cite{Ring}:
\begin{equation}
\hat{P}^{I}_{MK}=\frac{2I+1}{8\pi^{2}}\int d\Omega D^{I*}_{MK}(\Omega)\hat{R}(\Omega),
\end{equation}
for the angular momentum projection, and
\begin{equation}
\hat{P}^{N}=\frac{1}{2\pi}\int^{2\pi}_{0} d\phi \exp[-i\phi(\hat{N}-N)],
\end{equation}
for the particle number projection. It has been shown in Refs.~\cite{Doe.98,MER.01} that the particle number 
projection may cause troubles in the case that the exchange terms of the interaction are neglected.  For this
reason in our calculations we will not neglect any term.

Minimisation of the energy with respect
to the coefficients $f^{\sigma,I}_{\rho}(\beta)$ leads to  the Hill-Wheeler (HW) equation  \cite{HW}
\begin{equation}
\sum_{\rho'\beta'}\left(\mathcal{H}^{I}_{\rho\rho'}(\beta, \beta')-E^{\sigma I}\mathcal{N}^{I}_{\rho\rho'}(\beta, \beta')\right)f^{\sigma I}_{\rho'}(\beta')=0
\label{HW_eq}
\end{equation}
which has to be solved for each value of the angular momentum.  The GCM norm- and energy-overlaps  have been defined as:
\begin{eqnarray}
\mathcal{N}^{I}_{\rho \rho'}(\beta,\beta')&\equiv&\langle \Phi_{\rho} (\beta) | \hat{P}^{I}_{MK}\hat{P}^{N}\hat{P}^{Z}| \Phi_{\rho'}(\beta')\rangle\nonumber\\
\mathcal{H}^{I}_{\rho \rho'}(\beta,\beta')&\equiv&\langle \Phi_{\rho} (\beta) |H \hat{P}^{I}_{MK}\hat{P}^{N}\hat{P}^{Z}| \Phi_{\rho'}(\beta')\rangle.
\label{gcm_overlaps}
\end{eqnarray}

To cope with the problem of the linear dependence one first introduces an orthonormal basis defined by the eigenvalues $n^{\kappa I}$ and eigenvectors $u^{\kappa I}_{\rho}(\beta)$ of the norm overlap:
\begin{equation}
\sum_{\beta'\rho'}\mathcal{N}^{I}_{\rho \rho'} (\beta,\beta')u^{\kappa I}_{\rho'}(\beta')=n^{\kappa I}_{}u^{\kappa I}_{\rho}(\beta).
\end{equation}
This orthonormal basis is known as the natural basis and,  for $n^{\kappa I}$ values such that  $n^{\kappa I}/n^{I}_{max}>\zeta$, the  natural states are defined by:
\begin{equation}
|\kappa^{IM}\rangle=\sum_{\beta\rho}\frac{u^{\kappa I}_{\rho}(\beta)}{\sqrt{n^{\kappa I}}}\hat{P}^{I}_{MK}\hat{P}^{N}\hat{P}^{Z}| \Phi_{\rho}(\beta)\rangle.
\label{natstates}
\end{equation}
Obviously, a cutoff  $\zeta$ has to be introduced in the value  of the norm eigenvalues to avoid linear dependences \cite{RingAMP_Rel_09}.
Then, the HW equation is transformed into a normal eigenvalue problem:
\begin{equation}\label{HW_nat}
\sum_{\kappa'}\langle\kappa^{I}|\hat{H}|\kappa'^{I}\rangle g^{\sigma I}_{\kappa'}=E^{\sigma I}g^{\sigma I}_{\kappa}.
\end{equation}
From the coefficients $g^{\sigma I}_{\kappa}$ we can define the so-called collective wave functions 
\begin{equation}
p^{\sigma I}_{\rho}(\beta)=\sum_{\kappa}g^{\sigma I}_{\kappa}u^{\kappa I}_{\rho}(\beta)
\label{coll_wf}
\end{equation} 
that satisfy 
\begin{equation}
\sum_{\rho \beta} |p^{\sigma I}_{\rho}(\beta)|^2=1,  \;\;\; \forall \sigma,
\label{norm_coll_wf}
\end{equation}
and are equivalent to a density of probability.

Concerning the number of two-quasiparticle states considered in the GCM ansatz of Eq.~\ref{eqwf},  in this calculation we set the following energy cutoff condition:
\begin{equation}
E_{i}(\beta)+E_{j}(\beta)+E_{0}(\beta)\leq E(\beta_{min})+3.5 {\rm MeV}
\end{equation}
where $E_{i}(\beta)$ and $E_{j}(\beta)$ represent the quasiparticle energies of the states  $i$ and $j$.  $E_{0}(\beta)$ is the energy of the HFB state with deformation $\beta$ given by
\begin{equation}
E_0(\beta) = \langle \Phi_{0}(\beta)| \hat{H} | \Phi_{0}(\beta)\rangle,
\label{E_ground}
\end{equation}
with $| \Phi_{0}(\beta)\rangle$ the solution of Eq.~\ref{HFB_Eq} and $\beta_{min}$ represents the deformation of the HFB minimum.
$E(\beta_{min})$ is the energy minimum of the potential energy as a function of $\beta$. 
The convergence with this cutoff has been checked and it is very good. 

The Hamiltonian of Eq.~\ref{Hamiltonian} has been used extensively in many PSM calculations with great success \cite{PSMreview}.  We use the following strengths:  the monopole pairing strength is $G_{M}=(18.75\mp13.00(N-Z)/A)/A$, with the sign $"-"$ for neutrons and $"+"$ for protons and  the quadrupole pairing strength is $G_{Q}=0.16G_{M}$.  The quadrupole-quadrupole strength cannot be taken from the PSM because in that approach it is fixed by the experimentally measured $\beta$ deformation and it must be adjusted separately for each nucleus, see \cite{PSMreview}. 
For the strength of the quadrupole force, we take  $\chi_{\tau\tau^{\prime}}=\chi\alpha_{\tau}\alpha_{\tau^{\prime}}$. Here $\tau$ and $\tau^{\prime}$ can be either protons or neutrons. For protons $\alpha_{\pi}=(2Z/A)^{1/3}$ and for neutrons $\alpha_{\nu}=(2N/A)^{1/3}$. Furthermore,
$\chi=\chi^{\prime}/ b^{4}$ with $\chi^{\prime}=70A^{-1.4}$ MeV  and $b$ is the oscillator length ($b=A^{1/6}$). The choice of the quadrupole strength is inspired by the work of Baranger and Kumar, Ref.~\cite{BK1968-529}. These authors 
performed a very good analysis of the Pairing plus Quadrupole interaction \cite{BK}.  They  carried out very  extended calculations along the nuclide chart within different approaches.  In particular they calculated
ground states properties and potential energy surfaces in the HFB approach as well as in the beyond mean field
approach  in the framework of the
Bohr collective Hamiltonian  \cite{BK_coll}. All these facts have contributed to the wide acceptance of the Baranger Kumar approach. Unfortunately this model is not appropriate for some beyond mean field approaches since the single particle model space used by Baranger and Kumar consist of only two mayor shells and  it requires the use of a core to obtain good experimental agreement. The core provides, among others,  a contribution to the moment of inertia which is impossible to combine with an angular momentum projection.   As discussed in  Ref.~\cite{PSMreview} with a single particle space of three shells there is no need to introduce a core. The effective charge for calculating the E2 transitions are $1+1.5Z/A$ for protons and $1.5Z/A$ for neutrons as in Ref.~\cite{BK1968-529}.

In our calculations three major shells are used as the single particle space.  For rare earth nuclei they are $N=4,5,6$ for neutrons and $N=3,4,5$ for protons  \cite{PSMreview}.
For the single-particle energies we use the $\kappa$ and $\mu$ Nilsson parameters of Ref.~\cite{BR.85}  
for neutron and protons. Furthermore, the i$_{13/2}$  orbit in the $N=6$ shell for neutrons is shifted by $-0.025 \; \hbar\omega_0$. The h$_{11/2}$ orbit in the $N=5$ shell for protons is shifted by $-0.01\; \hbar\omega_0$, and the rest orbits in the same shell are shifted by $+0.10 \;\hbar\omega_0$.
 The purpose of these modifications of the single-particle energies is to obtain reasonable energy curves for either soft or well deformed Er isotopes. Without these modifications, the energy curves for $^{156}$Er will be extremely soft and obviously unreasonable.
It should be noticed that we do not determine our parameters by fitting the observed excitation energies of the $0^{+}_{2}$ states, as in other calculations \cite{IBM2004,IBM2012,Bohr2011,Bohr2013}. 

\section{Results and discussions}\label{sect3}

\begin{figure}[htbp]
\includegraphics[width=0.5\textwidth]{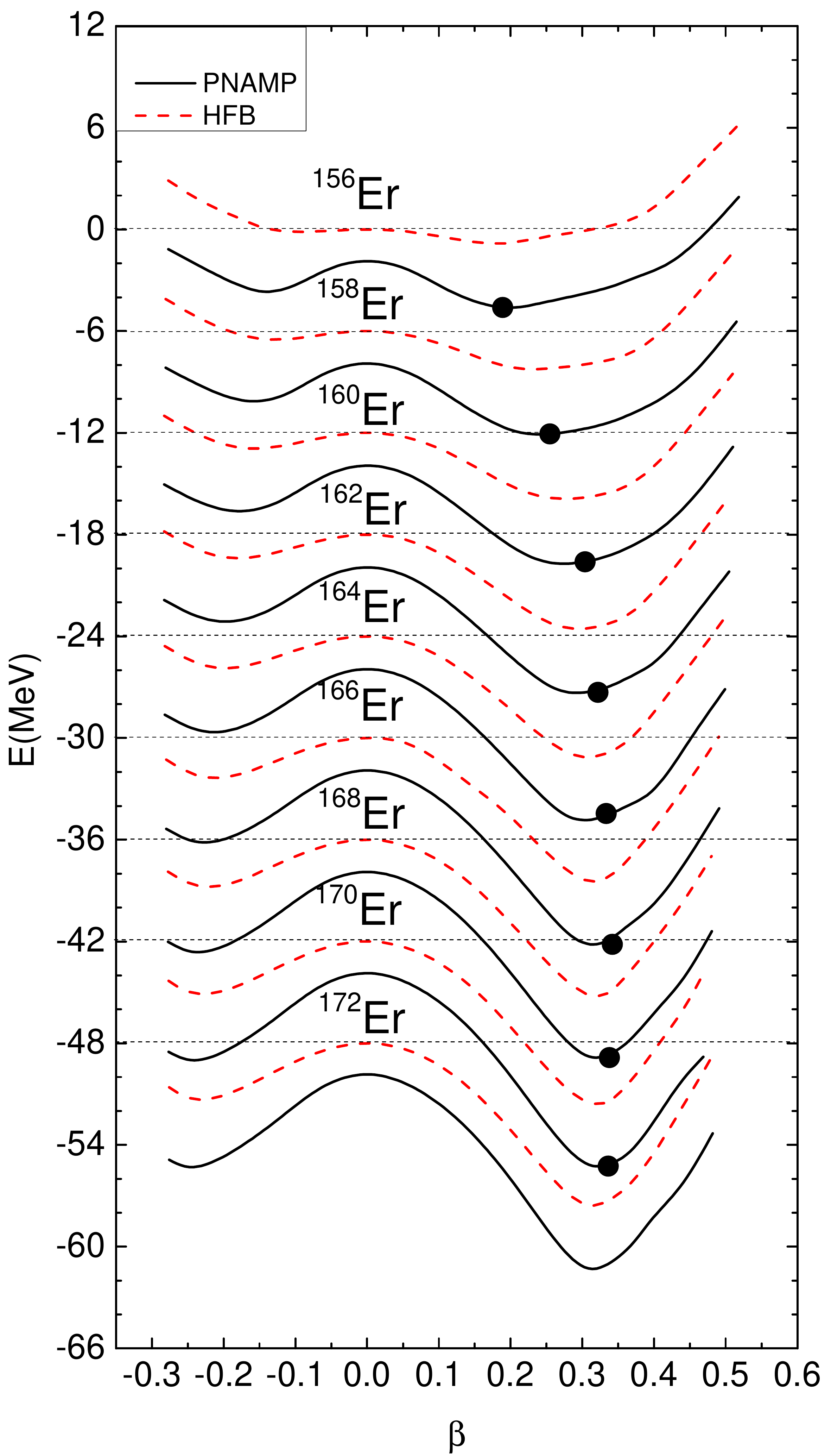}
\caption{ (Color online) Energy curves for $^{156-172}$Er calculated in the HFB approach, dashed lines,  and in the PNAMP one, continuous lines. The bullet on each curve represents the experimental value
of the deformation  \cite{beta_exp}.}
\label{Erpes}
\end{figure}

The Er isotopes studied in this work range from the soft ones (with neutron number $N=88$ or $90$) to the well deformed ones (up to $N=104$).  A very useful information is provided by the potential energy as a function of the deformation parameter $\beta$. The HFB ground state energy as a function of the deformation parameter $\beta$ is provided by Eq.~\ref{E_ground}.  The corresponding energy curves for the different Er isotopes  are shown in Fig.~\ref{Erpes} as dashed lines. The HFB energies  have been shifted in such a way that the energy  of the $\beta=0$ value of each nucleus coincide with successive  multiples of 6 MeV. The horizontal dotted lines  are plotted as an energy reference. In Fig.~\ref{Erpes} the transition from soft to rigid deformation is clearly visible. For $^{156}$Er the energy curve is very soft with two shallow coexisting minima at the prolate and oblate side. In this case one could expect large shape fluctuations and the excitation energy of the shape vibrational state should be relatively low. With increasing neutron number prolate and oblate minima develop around $|\beta| \approx 0.2-0.3$, with growing $\beta$ as N increases and a clear predominance of the prolate over the oblate shape. Large changes are found up to  $N=98$ ($^{166}$Er), for larger $N$ the potential energy curves look rather similar.  
 For the very well deformed isotopes one would expect small shape fluctuations around the prolate minimum, and the genuine shape vibrational excitations should appear at higher excitation energies. These curves qualitatively agree
 with the energy curves obtained with the Baranger and Kumar calculations of Ref.~\cite{BK1968-529}.    In the same figures we also plot the particle number and angular momentum projected (PNAMP) energies defined by
 \begin{equation}
 E^{I}(\beta) = \frac{\langle  \Phi_0(\beta)|\hat{H} \hat{P}^{I}\hat{P}^{N}\hat{P}^{Z}| \Phi_0(\beta)\rangle}{\langle  \Phi_0(\beta)| \hat{P}^{I}\hat{P}^{N}\hat{P}^{Z}| \Phi_0(\beta)\rangle},
\end{equation}
where we have introduced $\hat{P}^{I}=\hat{P}^{I}_{00}$. The PNAMP energies are relative to the corresponding HFB values.  The about 2 MeV energy gain obtained in the PNAMP approach with respect to the HFB one at  $\beta=0$ is due only to the particle number projection. At larger deformation we obtain about 4 MeV energy gain from both projections. The small energy gain  obtained from the angular momentum projection at very small deformations as compared with the larger ones has as a consequence that the prolate and the oblate minima are deeper than in the HFB approach. 
We have also plotted the values of the experimental deformations, where available \cite{beta_exp}, on the projected curves. As one can see these values are located very close to the absolute minimum of the corresponding energy curves.

\begin{figure}[htbp]
\centering
\includegraphics[width=0.5\textwidth]{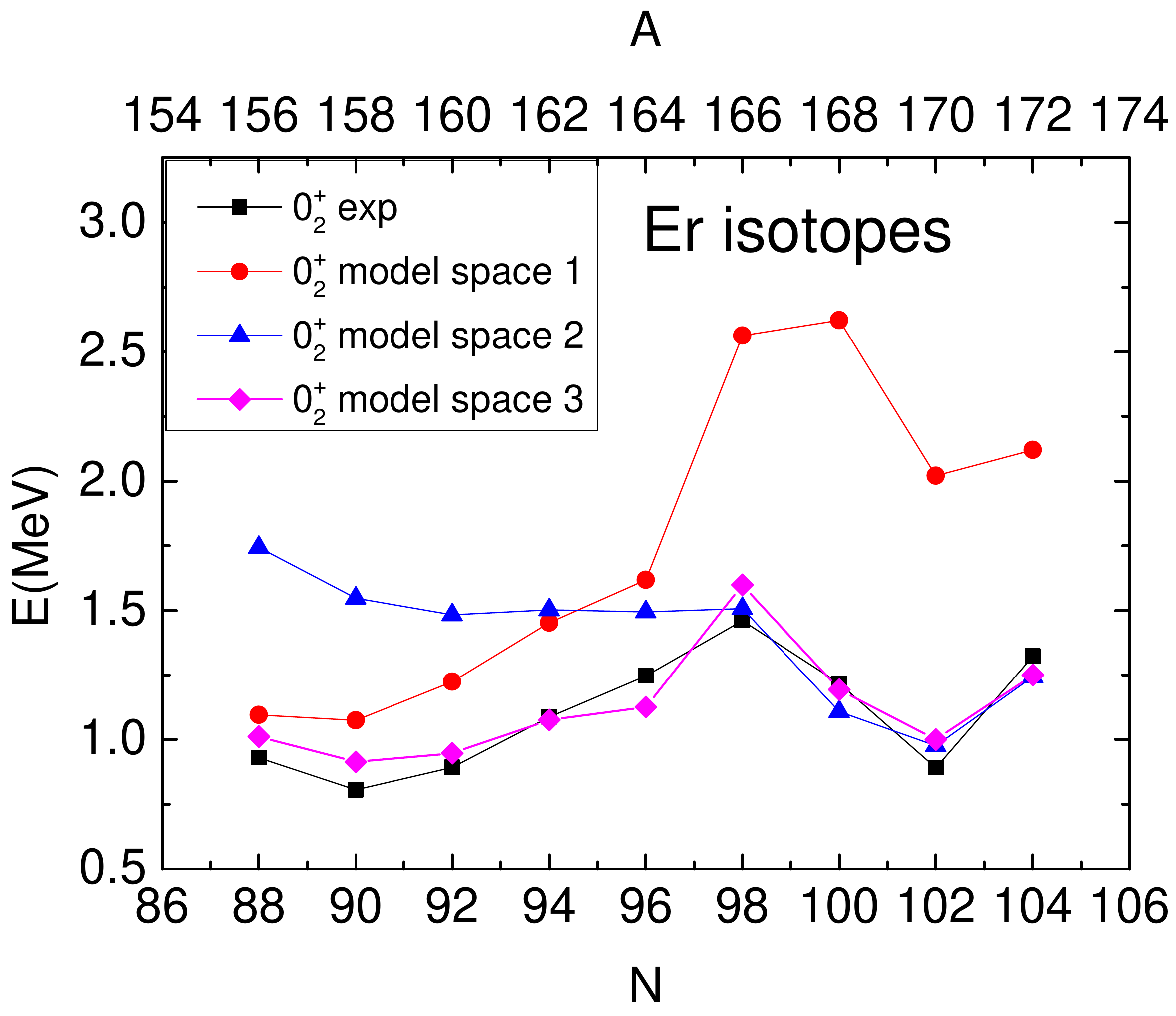}
\caption{ (Color online) $0^{+}_{2}$ energies calculated by three different model spaces. The model spaces 1, 2 and 3 correspond to the wave functions of Eq.~\ref{eqwf1}, Eq.~\ref{eqwf2} and Eq.~\ref{eqwf}, respectively. The experimental data \cite{ENSDF} is also shown for comparison.}
\label{GCMvs2QP}
\end{figure}

The next step is the solution of the Hill-Wheeler equation, Eq.~\ref{HW_nat}, which provides the energies and wave functions of the ground and excited states. In Fig.~\ref{GCMvs2QP} the experimental values for the excitation energies of the $0^{+}_{2}$ states for the Erbium isotopes are plotted (black squares). We observe an
increasing behavior of the excitation energies with the neutron number $N$ up to $N=98$, then a marked decrease up to $N=102$ and
again an increase towards $N=104$.  The rough structure of this behavior is understood in general terms: looking at a Nilsson diagram we find a deformed shell closure at $N=98$ and a less pronounced one at $N=104$. Since for shell closures one expects higher excitation energies, the mentioned closures roughly explain  the observed behavior.   In our model, see Eq.~\ref{eqwf}, we have two main degrees of freedom, namely, the shape vibrations and the two-quasiparticle excitations. To disentangle the different contributions we  present calculations in three different model spaces. The first model space includes HFB vacua of all deformations and does not include any two-quasiparticle state, i.e. the ansatz of Eq.~\ref{eqwf} is replaced by
\begin{eqnarray}\label{eqwf1}
|\sigma,IM \rangle&=&\int d\beta f^{\sigma,I}_{0}(\beta) \hat{P}^{I}_{MK}\hat{P}^{N}\hat{P}^{Z} |\Phi_{0}(\beta)\rangle.
\end{eqnarray}
The second model space includes the HFB vacuum corresponding to the minimum of the energy curve, as well as the two-quasiparticle states built on it, that means, 
\begin{eqnarray}\label{eqwf2}
|\sigma,IM \rangle&=& f^{\sigma,I}_{0}(\beta_{min}) \hat{P}^{I}_{MK}\hat{P}^{N}\hat{P}^{Z} |\Phi_{0}(\beta_{min})\rangle\nonumber\\
& & \;\;\;\;+\sum_{ij} f^{\sigma,I}_{ij}(\beta_{min}) \hat{P}^{I}_{MK}\hat{P}^{N}\hat{P}^{Z} |\Phi_{ij}(\beta_{min})\rangle\nonumber \\
& = & \sum_{\rho}  f^{\sigma,I}_{\rho}(\beta_{min}) \hat{P}^{I}_{MK}\hat{P}^{N}\hat{P}^{Z} |\Phi_{\rho}(\beta_{min})\rangle.
\end{eqnarray}
The third model space correspond to the full ansatz of Eq.~\ref{eqwf} and includes the HFB vacua of all deformations, as well as two-quasiparticle states built on each of them.
The results of the three calculations for the $0^{+}_{2}$ states  are plotted in Fig.~\ref{GCMvs2QP}:  red bullets for model space 1, blue triangles for model 2 and magenta diamonds for model 3.  We observe in models 1 and 2
a complementary behavior. While the $\beta$ vibrations provide better results for $N<98$, the two-quasiparticle
degree of freedom provides  better results for $N \geq 98$. These quantitative results reinforce what one qualitatively  would conclude from Fig.~\ref{Erpes}. Interestingly, the results of model 3 demonstrate how both degrees of freedom combine to reproduce the experimental results quite well. In a more careful analysis one observes that while for  $N=88, 90$ the shape fluctuations already provide rather good results and for $98\leq N \leq 104$ the two-quasiparticle states are sufficient, for $92\leq N \leq 96$  both degrees of freedom are clearly necessary.

\begin{figure}[htbp]
\includegraphics[width=0.5\textwidth]{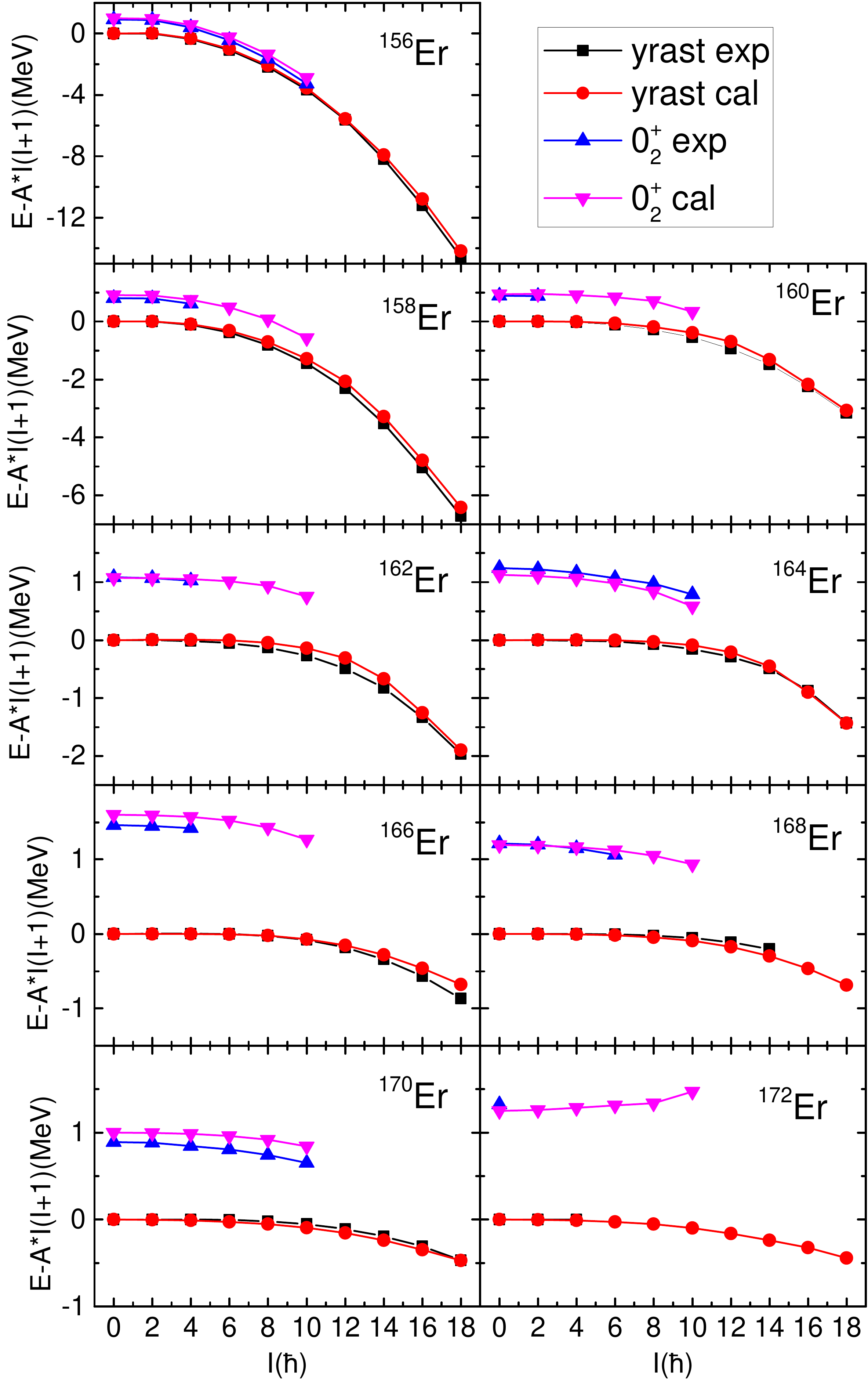}
\caption{(Color online) The calculated spectra of the yrast bands and the $0^{+}_{2}$ bands of $^{156-172}$Er, compared with the experimental data. The data are taken from \cite{ENSDF}.}
\label{Erspectrum}
\end{figure}

The eigenvalues of the Hill-Wheeler equation, Eq.~\ref{HW_eq},  with the full model space for different angular momenta are shown in Fig.~\ref{Erspectrum} for the Yrast band and the band built on the $0^{+}_{2}$ states (denoted as $0^{+}_{2}$ band in the following) for the Er isotopes. In the plots we have subtracted from the energies 
the rotational energy of a rigid rotor. The rotational constants have been fixed in such a way that the energy of the $2^{+}_{1}$ state is zero for each isotope.  The values of A in the plots are  $0.057, 032, 0.021, 0.017, 0.015$  ${\rm MeV/\hbar}^2$ for  $^{156-164}{\rm Er}$, respectively, and 0.013 for  $^{166-172}{\rm Er}$.
The levels have been classified into bands according to their transition probabilities.  Notice that the members of the $0^{+}_{2}$ bands do not coincide in general  with the lowest excited state with angular momentum $I$, i.e. with $I^{+}_{2}$ state.  For the $0^{+}_{2}$ bands and for very high angular momentum, the band structure may be dominated by two-quasiparticle and four-quasiparticle states. At present the four-quasiparticle configurations are not included in our model space, therefore we only show these bands up to $I=10\hbar$. 

\begin{figure}[htbp]
\centering
\includegraphics[width=0.5\textwidth]{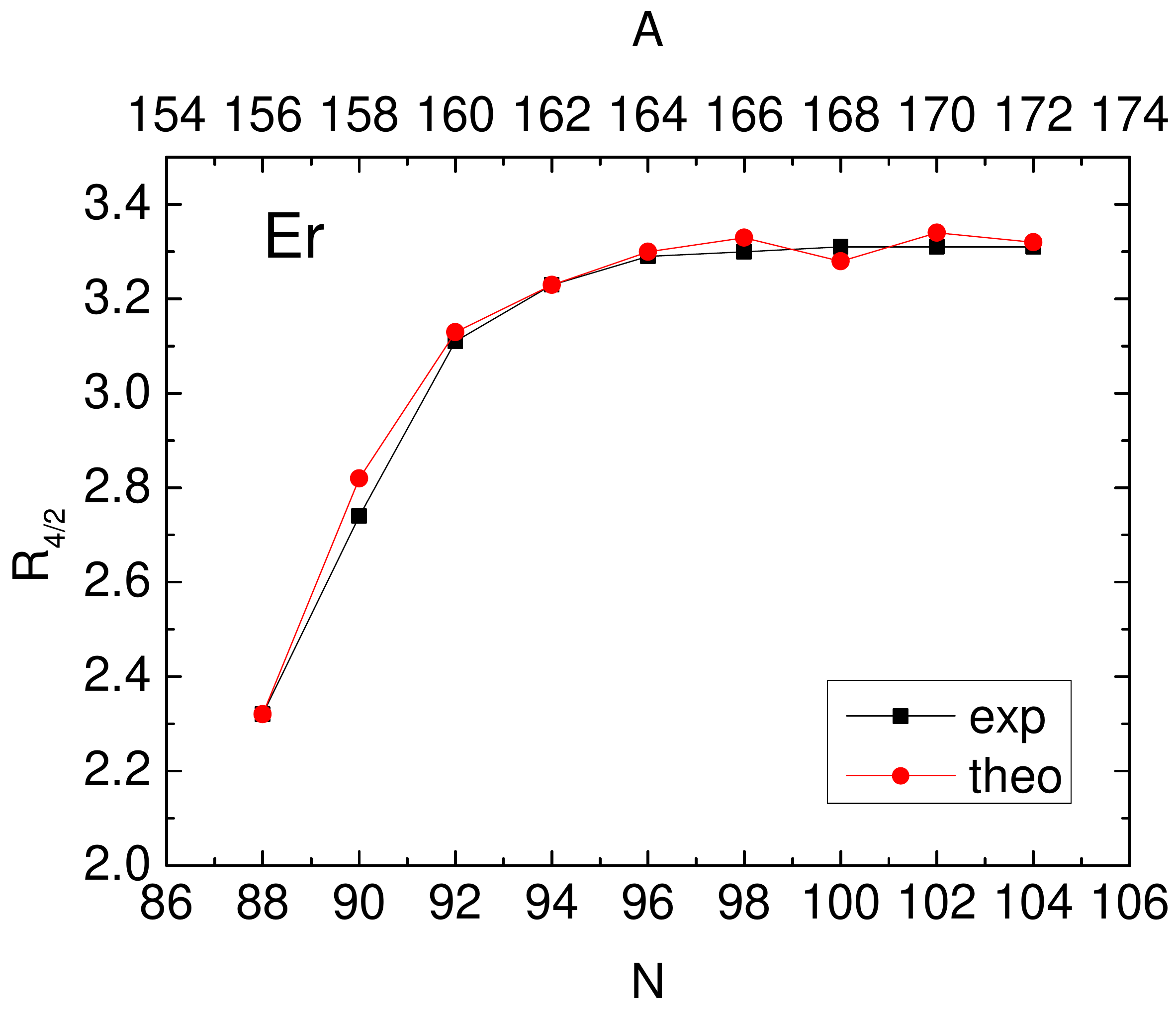}
\caption{ (Color online) The ratio $R_{4/2}=E(4^{+}_{1})/E(2^{+}_{1})$ for the Erbium isotopes.The data are taken from \cite{ENSDF}.}
\label{fig_4/2}
\end{figure}

\begin{figure}[htbp]
\centering
\includegraphics[width=0.5\textwidth]{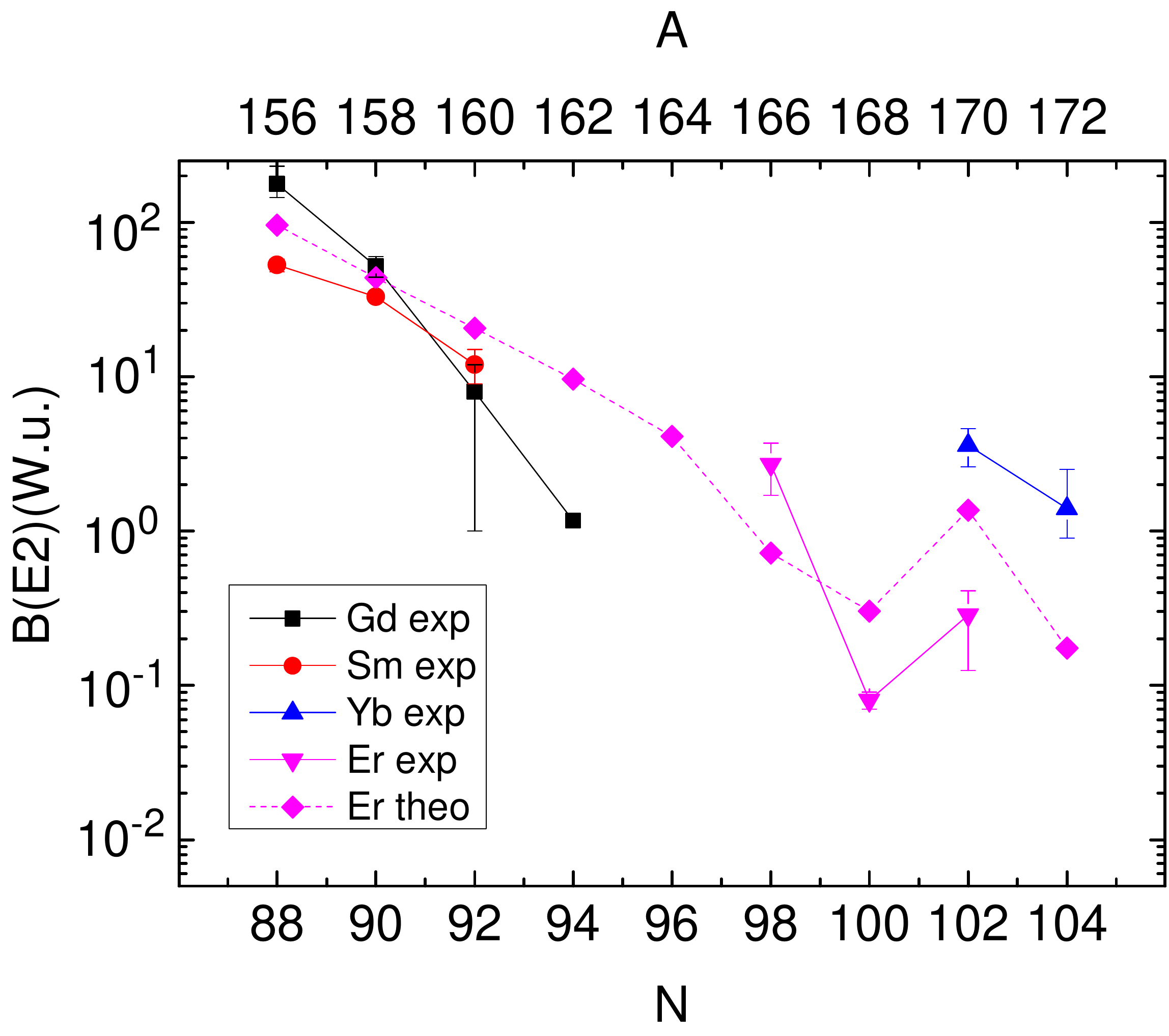}
\caption{ (Color online) Calculated $B(E2,0^{+}_{2}\rightarrow 2^{+}_{1})$ values for $^{156-172}$Er, compared with experimental data in neighbouring nuclei with the same neutron numbers. The data are  from \cite{ENSDF} except for $^{168}$Er and $^{170}$Er, which are taken from Ref.~\cite{Er168BE2} and Ref.~\cite{Er170BE2}, respectively.}
\label{ErBE2crossing}
\end{figure}

\begin{figure}[htbp]
	\centering
	\includegraphics[width=0.5\textwidth]{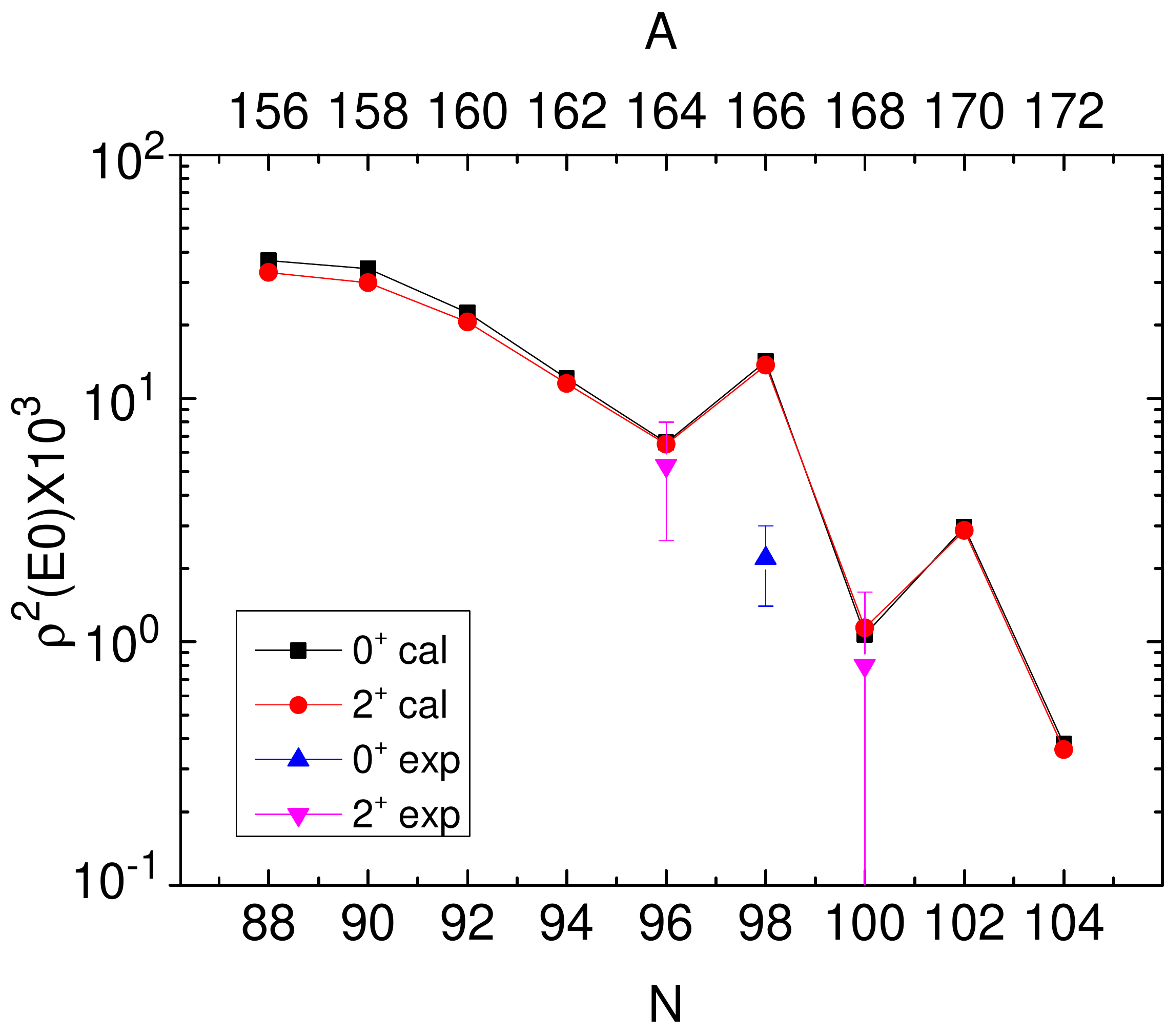}
	\caption{ (Color online) Calculated E0 transition probabilities for $^{156-172}$Er, compared with the experimental data where available. The data are taken from \cite{WZC.99}  for $^{164,168}$Er and from \cite{KS.05} for $^{166}$Er.}
	\label{E0}
\end{figure}

We observe that while the light Er isotopes deviate from a rigid rotor the heavier ones get close to it with increasing mass number. A similar behavior is observed for the excitation energy of the members of the $0^{+}_{2}$ band. Their excitation energies are very low for the light Erbium isotopes and higher for the heavier ones. This behavior is clearly related to the collectivity of the states. The members of the  $0^{+}_{2}$ band in the light Erbium isotopes are  collective whereas those in the heavier ones are less collective. These facts will show up in the transition probabilities, see Figs.~\ref{ErBE2crossing}-\ref{E0}.
In general we find a good agreement with the experimental data. In particular the good agreement  with the experiment for the low spin members of the Yrast bands indicates that the ratio $R_{4/2}=E(4^{+}_{1})/E(2^{+}_{1})$ is correctly reproduced. This ratio is plotted in Fig.~\ref{fig_4/2}.  It clearly reflects the transition from soft nuclei to well deformed ones. The description of this transition  is only possible because of shape mixing within the framework of the GCM. The higher spin members of the yrast bands are also well reproduced  by the inclusion of the two-quasiparticle states in the GCM ansatz. With these two-quasiparticle states we are able to describe band-crossing phenomenon at high spins. Although our main topic are the $0^{+}_{2}$ states, we can see the effect of the GCM as well as two-quasiparticle states in the systematic reproduction of the Yrast bands.

Not only the Yrast states are well described, the excitation energies of the $0^{+}_{2}$ bands are also reasonably well reproduced as shown in Fig.~\ref{Erspectrum}. This is an indication that our model space contains the important degrees of freedom for the bands based on the $0^{+}_{2}$ states. Many possible excitation modes have been proposed over the years for these $0^{+}$ excitations. Besides the $\beta$-vibrations and two-quasiparticle excitations mentioned in Sect.\ref{sect1}, there are other suggestions such as phonon excitation based on the $\gamma$-vibration \cite{IBM1994} or pairing isomers \cite{pairingisomer}. It is very difficult to include all of these possible degrees of freedom in a model space. Based on the good agreement with the experimental results it seems that in these Er isotopes, the important degrees of freedom are the shape vibration and the two-quasiparticle states. It should also be noticed that the moments of inertia of the $0^{+}_{2}$ bands are also well reproduced, especially in $^{164-168}$Er, in which the $0^{+}_{2}$ bands have larger moments of inertia than the Yrast bands.

\begin{figure}[htbp]
\centering
\includegraphics[width=0.48\textwidth]{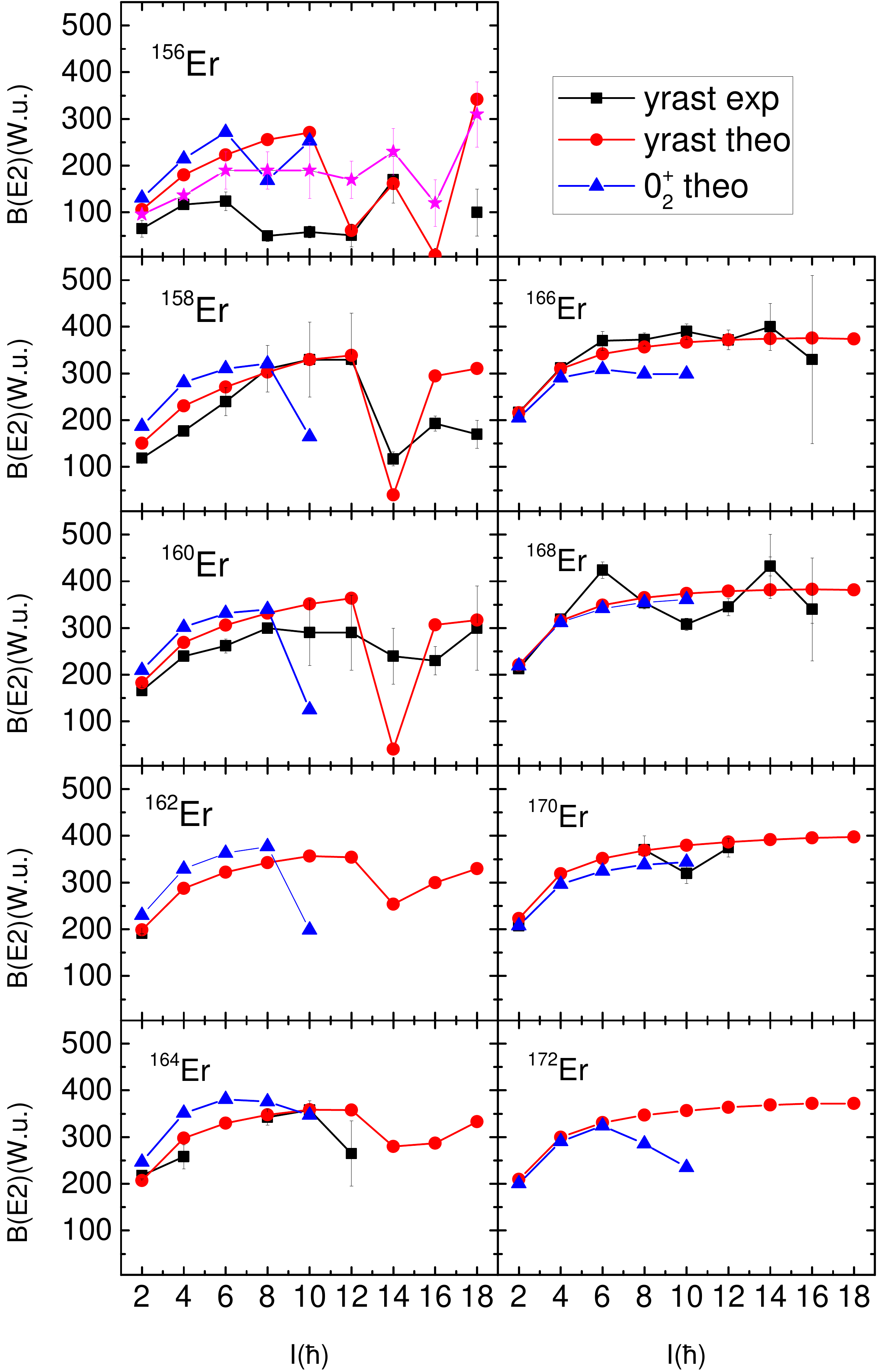}
\caption{(Color online) Calculated $B(E2,I\rightarrow I-2)$ values for the yrast bands and the $0^{+}_{2}$ bands of $^{156-172}$Er, compared with available data. The data are taken from \cite{ENSDF}.}
\label{ErBE2}
\end{figure}

Another relevant information is provided by  the intraband E2 transition probabilities. In Fig.~\ref{ErBE2}  we plot the calculated $B(E2,I\rightarrow I-2)$ values along the Yrast and the  $0^{+}_{2}$ bands  together with the available experimental information.  We first discuss the Yrast bands.  The calculated results are in reasonable agreement with the data except for $^{156}$Er.  In particular, the transition  probability from the
 $2^{+}_{1}$  to the  $0^{+}_{1}$ state increases roughly with the neutron number, corresponding to the transition towards the well deformed region. The decrease of the experimental $B(E2)$ values at high spin, which is a result of the band crossing,  is qualitatively reproduced by the theoretical results. The calculated intraband $B(E2)$ for $^{156}$Er disagree with the observed data. However, they are in qualitative agreement with the measured intraband $B(E2)$ in $^{154}$Dy, which is an isotone of $^{156}$Er. This may indicate that the disagreement of the intraband $B(E2)$ in $^{156}$Er does not mean a general failure of our model for soft nuclei. It may be related to some unknown reason in $^{156}$Er, which is not taken into account in our model. It might be suggested that $^{156}$Er is $\gamma$-unstable and we should include the $\gamma$ degree of freedom to describe the intraband $B(E2)$.
 
 The theoretical values for the intraband $B(E2)$ along the $0^{+}_{2}$ bands are also shown in the  Fig.~\ref{ErBE2} . They are of the same order as those of the Yrast bands. The decrease at around $I=10\hbar$ is due to the band crossing with a two-quasiparticle band.

Besides the intraband E2 transitions, the interband ones taking place between the $0^{+}_{2}$ and the $2^{+}_{1}$ states are also important. These  transition rates are related to the quadrupole collectivity of the initial $0^{+}$ state. Therefore to see the performance of our model it is important to compare the calculated interband $B(E2)$ values with experimental data. However, for the Er isotopes these transitions probabilities has been only measured for $^{166}$Er \cite{Er166dat}, $^{168}$Er \cite{Er168BE2} and $^{170}$Er \cite{Er170BE2}. In order to examine the neutron number dependence of this interband $B(E2)$, we also include the data measured in some Sm, Gd and Yb nuclei with the same neutron number as the Er isotopes.  The calculated interband $B(E2)$ values are shown in Fig.~\ref{ErBE2crossing} together with the data measured in Sm, Gd, Er and Yb isotopes. It is found that the interband $B(E2)$ values, if the $Z$ dependence is ignored, decrease with growing neutron number. This trend is reproduced by our calculations. For $^{156}$Er with $N=88$, the calculated interband $B(E2)$ is around 100 W.u., which is of the order estimated for a $\beta$-vibration \cite{Garrett} (although the estimation is made with the assumption of well deformed nuclei). This suggests that the $0^{+}_{2}$ state in this nuclei may be dominated by shape vibration. On the other hand, the calculated interband $B(E2)$ for $^{168}$Er is less than 0.4 W.u., which suggests that in this case the $0^{+}_{2}$ state may be dominated by two-quasiparticle excitations. These suggestions are in accordance with what was inferred from Fig.~\ref{GCMvs2QP}. Note that the variation of the interband $B(E2)$ within this nuclei is as large as three orders of magnitude, which indicates that these $0^{+}$ excitations are of very different structure. This large variation range has been well reproduced by our calculation, which is another justification of our GCM+2QP model space.

Another important observable are E0 transitions. They provide  important information about the mixing content of wave functions \cite{WHN.92}. In the absence of mixing one expects pure configurations and thereby small matrix elements and larger ones in the presence of mixing. The matrix element of the monopole transition  $m(E0)$  from the state $I^{+}_{i}$  to the $I^{+}_{f}$ is given by
\begin{equation}
m(E0)= \langle I^{+}_{f} | \sum_{k=1}^{A} e^{eff.}_{k} r_{k}^{2}|I^{+}_{i}\rangle,
\end{equation}
with $e^{eff.}_{k} =e(1+1.5Z/A)$ for protons and $e^{eff.}_{k} =e(1.5Z/A)$ for neutrons.
The states $I^{+}_{i}$  and $I^{+}_{f}$ are given by Eq.~\ref{eqwf}.
The E0 strength is usually measured by the dimensionless quantity $\rho^{2}(E0)$ related to $m(E0)$ by
\begin{equation}
\rho^{2}(E0) = \left| \frac{m(E0)}{eR_0^{2}}\right|^{2},
\end{equation}
where $R_0=r_0 A^{1/3}$  is the nuclear radius and $r_{0}=1.2$ fm.
In Fig.~\ref{E0} we display the $\rho^{2}(E0)$ values for the transitions from the first two members (i.e., $I=0^+ $ and $2^{+})$ of the $0^{+}_2$ band to the corresponding states of the ground band for the Erbium isotopes together with the experimental values. The  theoretical values for $I=2^+$, as expected, are very similar to the  $I=0^+$ ones.  We observe an exponential decrease of the strength with increasing mass number as in the previous case of $B(E2, 0^{+}_{2}\longrightarrow  2^{+}_{1} )$ displayed in Fig.~\ref{ErBE2crossing}. This behaviour clearly reflects the different stages of collectivity of the members of the $0^{+}_{2}$ band. The agreement with the experimental results for $^{164}$Er and  $^{168}$Er is good whereas for $^{166}$Er we
obtain a peak which is not observed experimentally. This is probably related to the change of the most relevant two-quasiparticle configuration from neutrons in  $^{164}$Er to protons in  $^{166}$Er, see Table~\ref{tab}.

\begin{figure}[htbp]
	\centering
	\includegraphics[width=0.5\textwidth]{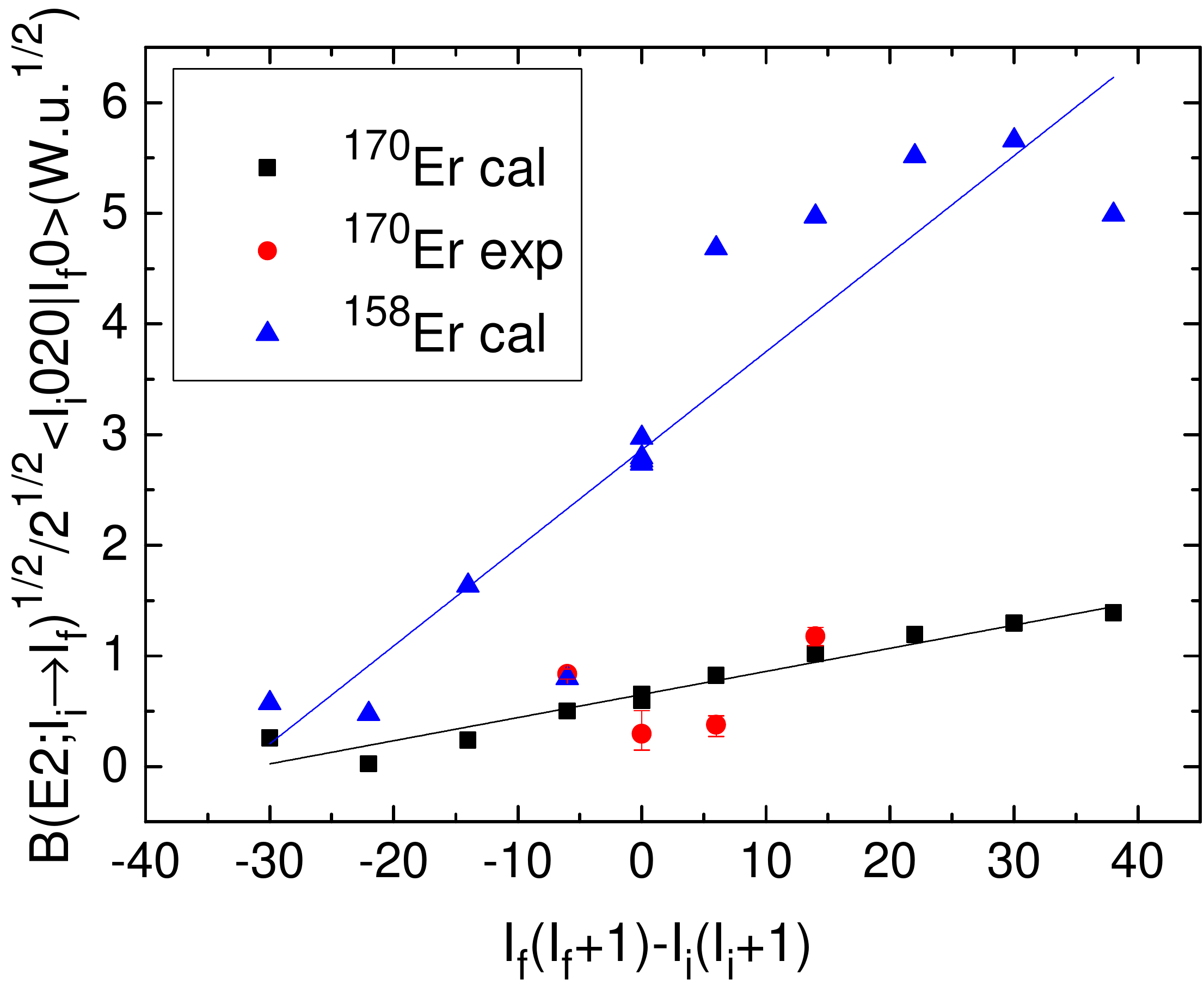}
	\caption{ (Color online) Mikhailov plot for the nucleus $^{170}$Er and  $^{158}$Er for transitions from the $0^{+}_{2}$ band to the ground band compared with the experimental data where available. The data are taken from \cite{ENSDF}.}
	\label{Mikhailov}
\end{figure}

Now we want to analyse the mixing of the unperturbed $0^{+}_{2}$ band and the ground band in the rotor model.   The simplest model for mixing of rotational bands was provided by Mikhailov for well deformed nuclei \cite{Mi.64}.  This model assumes that the origin of the mixing of the two bands is the Coriolis force. Assuming that the intrinsic quadrupole moment of the bands is the same one obtains \cite{casten } the simple formula 
\begin{equation}
\frac{\sqrt{B(E2, I^{E}_{i}\longrightarrow I^{g}_{f})}}{\sqrt{2}(I^{E}_{i}020|I^{g}_{f}0)}= M_1 - M_2\left[ I^{g}_{f}(I^{g}_{f}+1)-I^{E}_{i}(I^{E}_{i}+1) \right],
\label{eq:Mik}
\end{equation}
the superscript  $E$ denotes  the excited band, in our case the $0^{+}_{2}$ band and $g$  the ground band. If the assumptions of the model are fulfilled  $M_1$ and $M_2$ are
constants \cite{casten } and the plot, the so-called Mikhailov plot,  of the left hand side of Eq.~\ref{eq:Mik} as a function of $I^{g}_{f}(I^{g}_{f}+1)-I^{E}_{i}(I^{E}_{i}+1)$ should be a straight line. Notice that for each angular momentum $I_i$ the final states can have $I_f= I_i, I_i \pm 2$.   An isotope that satisfies the requirements of the model and for which there are some experimental results is $^{170}$Er.  In Fig.~\ref{Mikhailov} we represent the calculated transition probabilities together with the experimental data. We obtain a qualitative agreement with the experimental data and the fit of a straight line to the calculated points is a  good approximation indicating the validity of the model for this nucleus.  In contrast we also display in the same figure the calculated values for the nucleus $^{158}$Er. In this case we find that the Mikhailov predictions are not satisfied  indicating that the mixing of the bands has a different origin and that the assumptions of the model are not satisfied, see below.

Another analysis often used in the analysis of vibrational bands \cite{Ri.69}  is performed by mixing unperturbed bands but without assuming any specific form for the perturbation, the mixing parameters being adjusted as to obtain a good fit to the experiment. We have performed such an analysis to our data in the terms of Ref.~\cite{Ri.69}  for the nuclei $^{170}$Er and  $^{158}$Er and the results, not quoted here, are similar to the one based on the Mikhailov plots.  For the description of $^{170}$Er it is enough to mix the unperturbed ground and the $K=0^{+}$ bands  whereas for $^{158}$Er further bands will be required.

\begin{figure}[htbp]
\centering
\includegraphics[width=0.5\textwidth]{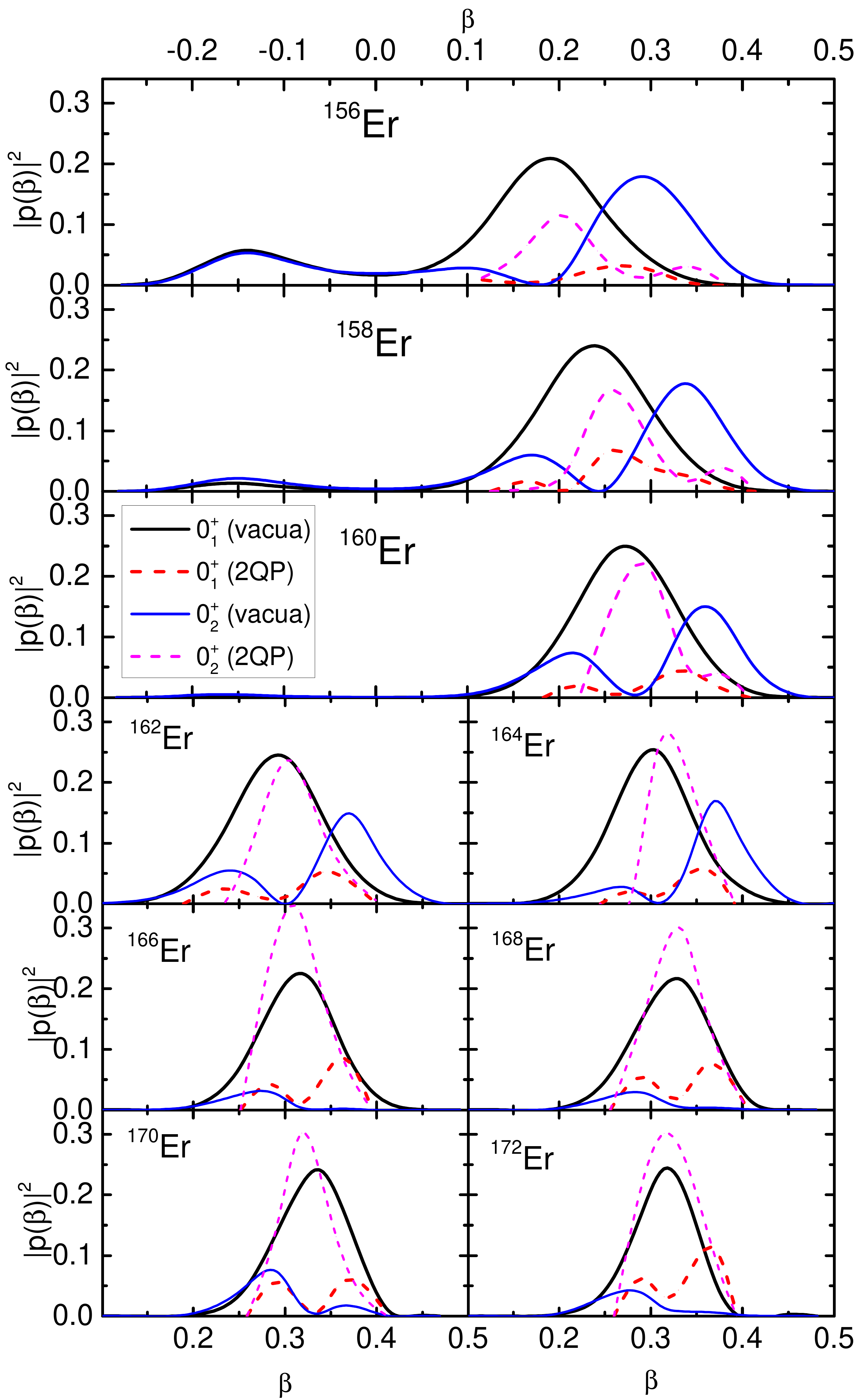}
\caption{ (Color online) Absolute squares of the collective wave functions calculated for the ground states and the $0^{+}_{2}$ states in $^{156-172}$Er. 
	 The  thick continuous black lines represent 
	 $|p^{\sigma I}_{0}(\beta)|^2$  and the thick dashed red lines $\sum_{\{ij\}} |p^{\sigma I}_{ij}(\beta)|^2$ as a function of the deformation parameter $\beta$ for the ground states. The   thin blue and magenta lines represent the corresponding amplitudes for the $0^{+}_{2}$ states.}
\label{Ercollective-2}
\end{figure}

In the  discussions above we compared the results of the calculations with the measured observables and demonstrated the reliability of our method. In the following we will focus on the collective wave functions.
According to Eq.~\ref{norm_coll_wf}, the normalization of the collective wave function Eq.~\ref{coll_wf} for a given state $\sigma$ and angular momentum $I$ can be written as 
\begin{equation} \label{coll_sep}
1= \sum_{\beta \rho}\big| p^{\sigma I}_{\rho}(\beta)\big|^2 =  \sum_{\beta}  \big|p^{\sigma I}_{\rho=0}(\beta)\big|^2 +
\sum_{\beta \rho,\rho\ne0}\big|p^{\sigma I}_{\rho}(\beta)\big|^2.
\end{equation}
The quantity $\big| p^{\sigma I}_{0}(\beta)\big|^2$ is related to the contribution of the $\beta$-constrained ground state (HFB vacuum), $|\Phi_0(\beta)\rangle$,  to the amplitude of the collective wave function, while 
 $\big| p^{\sigma I}_{\rho,\rho\ne0}(\beta)\big|^2  \equiv \big| p^{\sigma I}_{ij}(\beta)\big|^2$ is related to the amplitude of the
 two-quasiparticle state,  $\alpha^{\dagger}_{i} \alpha^{\dagger}_{i}|\Phi_0(\beta)\rangle$.  The separation of
 Eq.~\ref{coll_sep} makes sense because it allows to differentiate  the collective degrees from the single particle ones. 
We concentrate on the wave functions of the ground and first excited state, $\sigma =1 $ and $2$, respectively,  and for $I=0\;\hbar$.  In Fig.~\ref{Ercollective-2} we plot the quantities 
$|p^{\sigma I}_{0}(\beta)|^2$  and $\sum_{\{ij\}} |p^{\sigma I}_{ij}(\beta)|^2$ as a function of the deformation parameter $\beta$ for the different Erbium isotopes.   The HFB vacua contributions to the $0^{+}_{1}$ state are represented by thick black solid lines. In the top panels one observes that for the $^{156,158,160}$Er isotopes there is a contribution from the oblate part. For the heavier isotopes only the prolate part is different from zero.  Furthermore it can be seen that  the width of the amplitude distribution gets smaller as the neutron number increases corresponding  to the depth of the prolate well. This is in accordance with the conclusion drawn from the energy curves shown in Fig.~\ref{Erpes}. The contribution of the two-quasiparticle states to the wave function of the $0^{+}_{1}$ states, thick dashed red lines, is negligible for the oblate part  and rather small for the prolate part of all isotopes though it increases with growing neutron number. That means that for the ground states the relevance of the two-quasiparticle states is always small. The wave functions of the ground states are dominated by the HFB vacua, as one would expect.

The amplitudes for the $0^{+}_{2}$ states are also plotted in Fig.~\ref{Ercollective-2}.  The HFB vacua part is represented by thin solid blue lines and the two-quasiparticle part by short dashed thin magenta lines. 
Notice that the two-quasiparticle part of the $0^{+}_{2}$ states peaks close to the $\beta$-values where the  vacuum part of the  $0^{+}_{1}$ states does as it should be.
A first glance reveals that
now the two-quasiparticle parts play a much more important role than for the  $0^{+}_{1}$ states. As a matter of fact the HFB vacua contributions are larger than the two-quasiparticle part only for the  $^{156-162}$Er isotopes while for  $^{164-172}$Er the latter one predominates.  Interestingly the HFB vacua part contribution to the wave functions displays  a node structure \footnote{It may look that for $^{158-160}$Er the thin solid blue lines have two nodes, but it is just because some of the values around $\beta=0$ are very small and seems to vanish.  The real nodes corresponds to positive $\beta$ values.} for $^{156-164}$Er characteristic for shape vibrations \cite{Ring,GCM2013}. For $^{166,172}$Er the HFB vacua contributions are very small and do not show a nodal structure.   The picture that emerges from this analysis is that the $0^{+}_{2}$ states in the light Erbium isotopes are dominated by shape vibrations while in the heavier ones the two-quasiparticles excitations predominate. Though at first sight it seems a smooth change from one nucleus to the other,  small details of the wave function may cause large changes in some observables. Thus the increase in the $B(E2,0^{+}_{2}\rightarrow 2^{+}_{1})$ transition probability from $^{168}$Er to  $^{170}$Er, see Fig.~\ref{ErBE2crossing},  can be explained noticing that the HFB vacua contribution to the wave function of the $0^{+}_{1}$ state \footnote{ For a well deformed nucleus the deformation of the $2^{+}_{1}$ state is the same as the $0^{+}_{1}$.} peaks at higher deformation than the corresponding to $^{168}$Er.  
The span of several orders of magnitude observed in Fig.~\ref{ErBE2crossing} for the interband $B(E2)$ transition probabilities from the $0^{+}_{2}$ states is not surprising if one considers the large variety of wave functions .
\begin{figure}[htbp]
\centering
\includegraphics[width=0.5\textwidth]{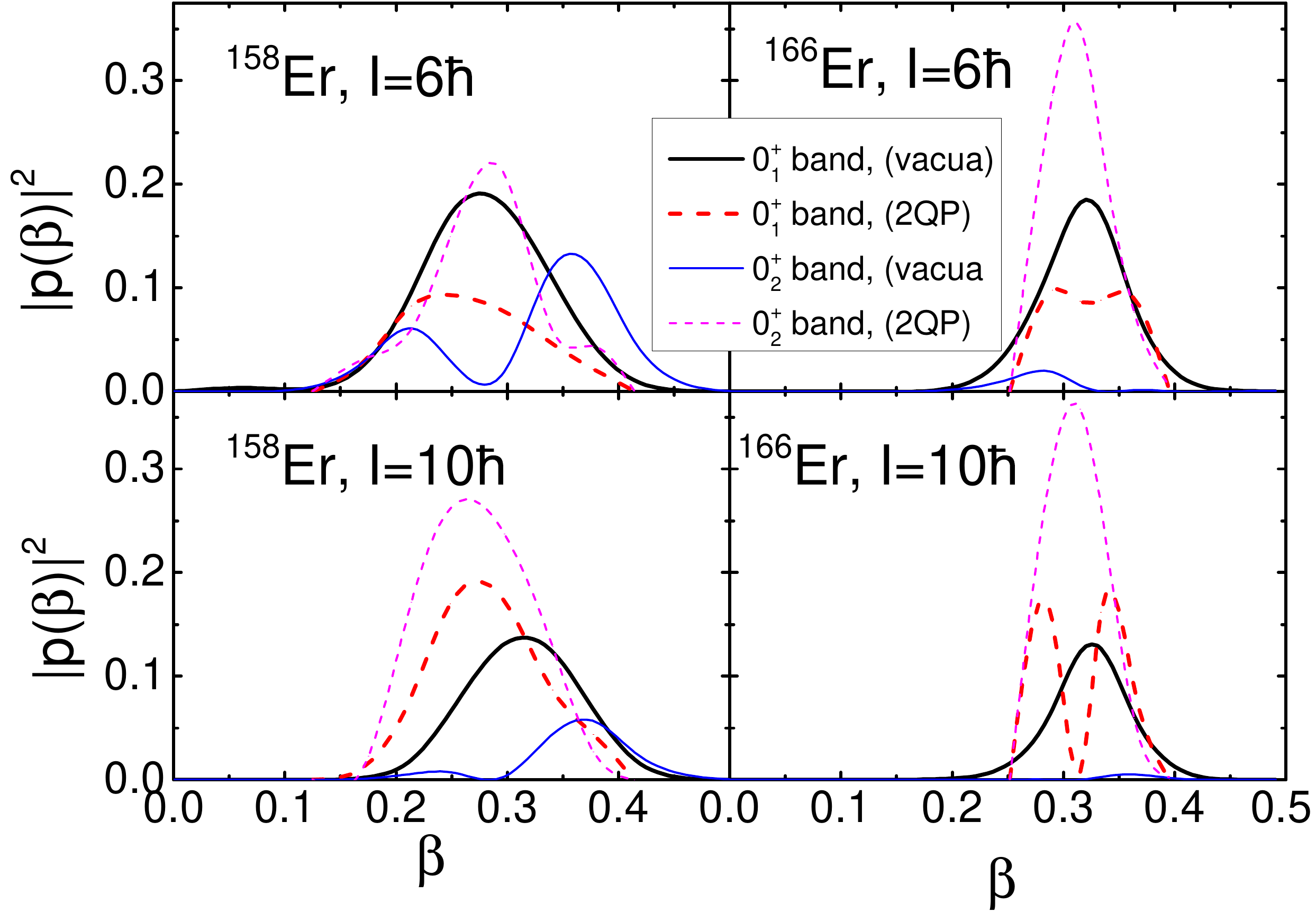}
\caption{ (Color online) Same as Fig.~\ref{Ercollective-2} but for angular momentum $6\;\hbar$, top panels, and $10\;\hbar$, lower panels,  and  two nuclei $^{158}$Er, left panels, and $^{166}$Er, right panels.}
\label{Ercollective-J}
\end{figure}

To know the evolution of the composition of the wave functions with the angular momentum we have plotted in Fig.~\ref{Ercollective-J} the same quantities as in Fig.~\ref{Ercollective-2} but for angular
momenta $6\;\hbar$ and $10\;\hbar$ and only for the nuclei $^{158}$Er and $^{166}$Er. For the soft 
$^{158}$Er and for the $I= 6\;\hbar$ member of the Yrast band we observe a decrease of the HFB vacua contribution, as compared with $I= 0\;\hbar$,  specially for small prolate deformations and in particular for the oblate part which  vanishes (not shown here).   The two-quasiparticle contribution now increases vigorously and its maximum is about half of the HFB vacua. Concerning the $I=6$ state of the $0^{+}_{2}$ band  we observe the same tendency but now the increase of the  two-quasiparticle part is even stronger than before. For $I=10\;\hbar$ and for both states the HFB vacua contribution diminishes even more and the  two-quasiparticle part increases. Furthermore for the Yrast state we observe a shift of the HFB vacua contribution to larger deformations and another of the two-quasiparticle part to smaller deformations. For $^{166}$Er, right panels, the role of the quasiparticles is more important than for
$^{158}$Er, in particular for the member of the $0^{+}_{2}$ band.

\begin{figure}[htbp]
\centering
\includegraphics[width=0.5\textwidth]{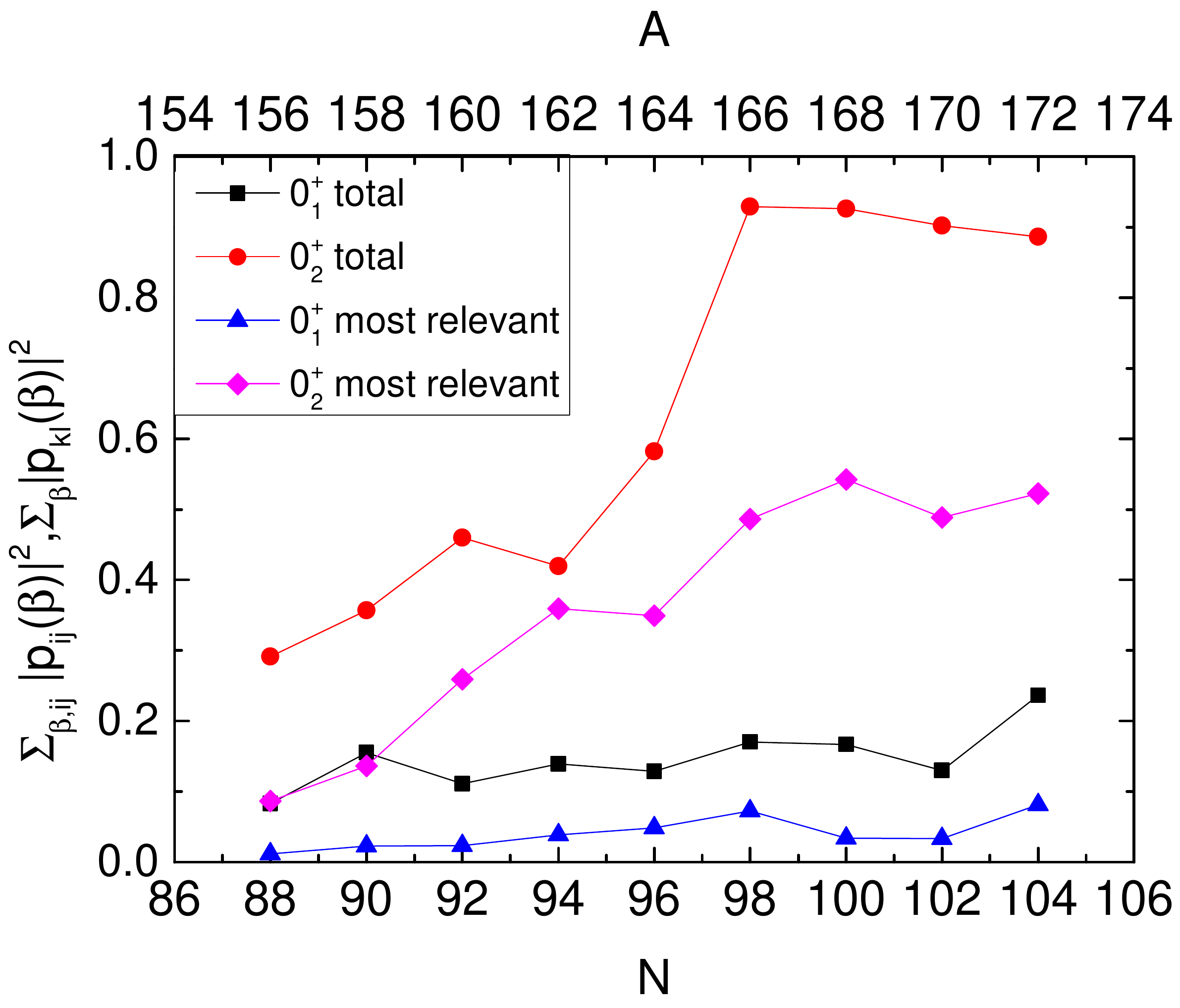}
\caption{ (Color online) Contribution of the two-quasiparticle states to the ground states and the $0^{+}_{2}$ states in $^{156-172}$Er, measured by the summation of the absolute square of the collective amplitudes over the two-quasiparticle states and all shapes. The contributions from the most relevant two-quasiparticle configurations are also shown.}
\label{Er2qpweight}
\end{figure}

In Fig.~\ref{Ercollective-2} we have presented the contribution of the two-quasiparticle states for each deformation $\beta$ for the Erbium isotopes. To evaluate the relevance of the two-quasiparticle excitations for a given nucleus we need the total contribution of the two-quasiparticle to all $\beta$'s, i.e., the last term of Eq.~\ref{coll_sep}. This term  is represented in Fig.~\ref{Er2qpweight} for the Er isotopes. The full squares are for the ground state and the bullets for the $0^{+}_{2}$ state. It is shown that the relevance of the two-quasiparticle excitation for the $0^{+}_{2}$ states increases with the neutron number up to $N=98$ and then remains more or less constant. In particular, the $0^{+}_{2}$ state  is to a large extent a two-quasiparticle excitation in the$^{166-172}$Er isotopes. 

We are also interested to know the indices  $(k,l)$ of the pair of quasiparticles  providing the largest  contribution to the sum just discussed. This contribution is given by $\sum_{\beta}  \big|p^{\sigma I}_{kl}(\beta)\big|^2$ and  plotted in Fig.~\ref{Er2qpweight} as triangles  and diamonds for the ground state and the $0^{+}_{2}$ state, respectively.  It is seen that the most relevant two-quasiparticle configuration accounts for more or less half of the total contribution of all two-quasiparticle configurations. Such a structure of the two-quasiparticle combinations is in qualitative accordance with the results of the QPM \cite{QPM-Er166,QPM-Er168,QPM-Gd158} and PSM \cite{PSM-Gd158,PSM-Er168} calculations. The Nilsson quantum numbers of the most relevant two-quasiparticle configurations are listed in Table.\ref{tab}. They are built by time-reversal conjugate states and are the lowest two-quasiparticle states. This is also in accordance with the QPM and PSM results.

\begin{table}[htbp]
\centering
\caption{\label{tab} The most relevant two-quasiparticle configurations in the $0^{+}_{2}$ states in $^{156-172}$Er.}
\begin{tabular}{cc}
  \hline
   & Two-quasiparticle configuration \\
  \hline
  $^{156}$Er & $\pi  \frac{7}{2}[404]\otimes   \pi \frac{7}{2}[404]$ \\
  $^{158}$Er & $\nu \frac{11}{2}[505]\otimes \nu\frac{11}{2}[505]$ \\
  $^{160}$Er & $\nu \frac{11}{2}[505]\otimes \nu\frac{11}{2}[505]$ \\
  $^{162}$Er & $\nu \frac{11}{2}[505]\otimes \nu\frac{11}{2}[505]$ \\
  $^{164}$Er & $\nu \frac{11}{2}[505]\otimes  \nu\frac{11}{2}[505]$ \\
  $^{166}$Er & $\pi  \frac{1}{2}[411]\otimes  \pi \frac{1}{2}[411]$ \\
  $^{168}$Er & $\nu \frac{1}{2}[521]\otimes \nu \frac{1}{2}[521]$ \\
  $^{170}$Er & $\nu \frac{5}{2}[512]\otimes \nu \frac{5}{2}[512]$ \\
  $^{172}$Er & $\nu \frac{7}{2}[514]\otimes \nu \frac{7}{2}[514]$ \\
  \hline
\end{tabular}
\end{table}

\section{Summary}\label{sect4}

In this work the generator coordinate method (GCM) is generalised by including two-quasiparticle excitations built on generating functions of different deformations as well as the ground states. In this way the collective vibrational shape fluctuations  and the non-collective quasiparticle excitation are taken into account within a common framework. With such a model space and the separable pairing plus quadrupole Hamiltonian,  calculations for Er isotopes with neutron number $N=88-104$ have been performed. The strengths of the pairing and quadrupole force used in the calculation are determined in a systematic way, and the single-particle energies are adjusted in order to get reasonable potential energy curves. The calculated spectra of the Yrast bands and the bands built on the $0^{+}_{2}$ states are in good agreement with experimental data. The intraband (along the Yrast bands) and interband (from the $0^{+}_{2}$ states to the $2^{+}_{1}$ states) E2 transitions are also reasonably well reproduced (except for the intraband $B(E2)$ in $^{156}$Er). The transition from soft to rigid deformation with the increasing neutron number is also well reproduced by our calculations.

The properties of the $0^{+}_{2}$ states are studied in detail. The good agreement between the calculated excitation energies of these $0^{+}$ states and the experimental data suggests that the two degrees of freedom, i.e. shape vibration and two-quasiparticle excitation, are sufficient for the description of these $0^{+}$ states. The large variation of the $B(E2,0^{+}_{2}\rightarrow2^{+}_{1})$ values, shown in our calculated results as well as in the measured experimental data for several neighbouring nuclei, indicates that the role played by these two degrees of freedom in the $0^{+}_{2}$ states are very different for different isotopes. Studies of the collective wave functions of these $0^{+}$ states show that in $^{156}$Er the $0^{+}_{2}$ state is dominated by shape vibration, while the two-particle excitation only plays a minor role. On the other hand, in $^{168}$Er the $0^{+}_{2}$ state is dominated by two-quasiparticle excitations, while the contribution from the shape vibration is almost negligible. The situation in other Er isotopes lie between these two extremes, in which both degrees of freedom are necessary for a reasonable description. Generally speaking, the relevance of two-quasiparticle excitations increase with the neutron number in the Er isotopes studied. The contribution of shape vibration decrease with the neutron number. The exception is $^{170}$Er in which the shape vibrations still make a larger contribution to the $0^{+}_{2}$ state as compared with $^{168}$Er.

One limitation of this calculation is that the axial symmetry is preserved and the $\gamma$-degree of freedom is not taken into account. As a consequence we do not have $\gamma$-vibrational states in our calculations. Therefore we could not study the E2 transition between the $0^{+}_{2}$ states to the $\gamma$-vibrational state, which is also an important quantity discussed in the studies of the nature of the $0^{+}_{2}$ states. Because of these we shall include the $\gamma$ degree of freedom in our future work.

We would lastly mention that a potential application of the theory would be the calculation of neutron pair transfer to investigate the $K=0^{+}$ bands \cite{Garrett}. Two neutron transfer in deformed nuclei is usually calculated in the HFB or QRPA  theories \cite{BHS.73,ER.87}. For deformed superfluid nuclei the particle number and the angular momentum are not conserved quantities and the use of a symmetry conserving theory like the one presented in this work may reveal important clues on the pair transfer phenomenon. 

\begin{acknowledgments}
Fang-Qi Chen gratefully acknowledges discussions with Prof. Yang Sun. This work was supported by the Spanish Ministerio de Econom\'ia y Competitividad under contracts FPA2011-29854-C04-04, FPA2014-57196-C5-2-P.

\end{acknowledgments}

\end{document}